\documentclass[reprint,amsmath,amssymb,aps,]{revtex4-2}
\usepackage{graphicx}
\usepackage{hyperref}
\hypersetup{%
    colorlinks=true
}
\begin{document}
\preprint{APS/123-QED}

\title{Inverse magnetocaloric effect in synthetic antiferromagnets}

\author{D.\ M.\ Polishchuk}
\altaffiliation[Also at ]{Institute of Magnetism, NASU/MESU, 03142 Kyiv, Ukraine}
\email{dpol@kth.se.}
\author{M.\ Persson}
\affiliation{Nanostructure Physics, Royal Institute of Technology, 10691 Stockholm, Sweden} 
\author{M.\ M.\ Kulyk}
\altaffiliation[Also at ]{Institute of Physics, NASU, 03028 Kyiv, Ukraine.}
\author{E.\ Holmgren}
\author{G.\ Pasquale}
\author{V.\ Korenivski}
\affiliation{Nanostructure Physics, Royal Institute of Technology, 10691 Stockholm, Sweden}
 
\author{A.\ Ullah}
\author{R.\ Skomski}
\affiliation{Department of Physics and Astronomy and the Nebraska Center for Materials and Nanoscience, University of Nebraska, 68588-0111 Lincoln, Nebraska, USA}

\begin{abstract}
The magnetocaloric effect in exchange-coupled synthetic-antiferromagnet multilayers is investigated experimentally and theoretically. We observe a temperature-controlled inversion of the effect, where the entropy increases on switching the individual ferromagnetic layers from anti-parallel to parallel alignment near their Curie point. Using a microscopic analytical model as well as numerical atomistic-spin simulations of the system, we explain the observed effect as due to the interplay between the intra- and inter-layer exchange interactions, which either add up or counteract to effectively modulate the Curie temperature of the dilute ferromagnetic layers. The proposed method of designing tunable, strongly magneto-caloric materials should be of interest for such applications as heat-assisted spintronics and magnetic refrigeration.
\end{abstract}

\maketitle

The magnetocaloric effect (MCE) is a field-driven energy exchange between the magnon and phonon subsystems of a magnetic material, which manifests as an adiabatic change in its temperature~\cite{franco18,miller14}. The discovery of a giant MCE in Gd-based alloys at room temperature~\cite{pecharsky97PRL,pecharsky97APL} has shown the prospect for solid-state magnetic refrigeration, believed to be more environmentally friendly and energy-efficient than the conventional gas-based refrigeration~\cite{franco18,gschneidner05,gutfleisch10}. The giant MCE, later observed also in transition-metal~\cite{wada01,rong07,tegusNature02} and Heusler-type alloys~\cite{hu00,liu12}, is due to a field-induced spin rearrangement in the vicinity of the magnetic phase transition, which is often accompanied by a first-order structural phase transition~\cite{pecharsky03,tegus02,trung10}. Other novel approaches to enhancing the MCE are based on the finite-size effects in thin films and multilayers~\cite{barman20,miller14,dung20} and a family of spin-caloritronic effects in heavy-metal/ferromagnet based nanostructures~\cite{bauer12,boona14}.

Nanostructuring was recently shown to yield greatly enhanced MCE in materials containing no rare-earth elements~\cite{miller14,skomski08,mukherjee09,skomski10,michalski11,michalski12,polishchuk18} via controlling the relevant anisotropy and exchange parameters~\cite{sellmyer06}. The approach exploits the finite-size effects in nanostructures, where the interface and bulk contributions are energetically comparable. A classic example of a finite-size magnetic effect is the indirect exchange coupling in ferromagnetic/nonmagnetic (F/N) metallic multilayers known as the Ruderman-Kittel-Kasuya-Yosida (RKKY) interaction~\cite{grunberg86}. The strength and sign of RKKY can be tuned by varying the thickness of N~\cite{parkin90,parkin91}, which is used for designing such artificial materials as synthetic antiferromagnets~\cite{duine18}.

Optimizing interfacial exchange can lead to enhanced MCE properties in transition-metal nanostructures~\cite{skomski08,mukherjee09,skomski10,michalski11,michalski12}. A large MCE was predicted for a system of macrospins ($\sim$100~\textmu$_\text{B}$) embedded into a weakly magnetic matrix, amplified by the interparticle exchange~\cite{skomski08,skomski10}. Subsequent experiments on magnetic nano-composites~\cite{michalski11,michalski12} and RKKY-coupled superlattices~\cite{mukherjee09} have indeed shown large isothermal entropy changes of up to $\Delta S_{\text{max}} =–0.4$~J/kg~K in a field of 7~T. A conceptually different multilayer system, based on thermally controlled RKKY~\cite{polishchuk17PRB,polishchuk17EPL,dpol-apl20}, showed a giant MCE of $\Delta S_{\text{max}} =–1.4$~J/kg~K in fields as low as 10~mT~\cite{polishchuk18}. The magneto-calorically active layer was a dilute ferromagnet undergoing an RKKY-induced magnetic phase transition on antiparallel-to-parallel magnetization switching in the system. In all these studies the observed MCE was \textit{normal}, in which the applied magnetic field \textit{decreases} the system’s magnetic entropy ($\Delta S_{\text{max}}<0$). As we show in this work, nanostructuring can be used to invert the MCE, such that the magnetic entropy \textit{increases} on saturation. To date, \emph{inverse} MCE was observed only for some bulk materials, such as intermetallic compounds~\cite{tegus02,zhang04} and Heusler-type alloys~\cite{marcos03,krenke05}. 

In this Letter, we show that antiferromagnetic RKKY combined with tuning the Curie point of the ferromagnetic layers via dilution can be used to \textit{invert} the MCE in multilayers, such that the field-driven magnetization switching into the \textit{nominally} saturated parallel state \textit{increases} the magnetic entropy of the system ($\Delta S_{\text{max}} > 0$; in contrast to normal MCE with $\Delta S_{\text{max}} < 0$). The material system we demonstrate has a high degree of tuneability of the intra- and inter-layer exchange, undergoes a change of sign in its MCE as a function of temperature, with up to 25\% of the ferromagnetic component spin-disordered by field-switching the interlayer RKKY coupling in moderate fields ($\sim0.1$~T).

\begin{figure}[t]
\includegraphics{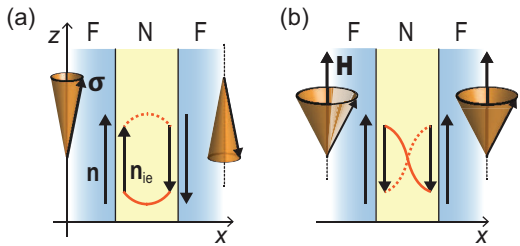}
\caption{\label{fig:effect} Effect of external field on spin alignment and entropy of thermally agitated antiferromagnetically RKKY-coupled F-N-F trilayer. Cone angle illustrates average thermal deviation of atomic spin directions for (a) zero-field AP and (b) saturating-field P configurations.}
\end{figure}

Figure 1 shows how the spin configuration of a F-N-F trilayer, in which the two ferromagnets coupled by antiferromagnetic RKKY-exchange and subject to thermal agitation, switch from antiparallel (AP) to parallel (P) alignment in an external field. Thermal excitations cause the atomic spin directions ${\bf\sigma}$ to fluctuate in each layer. The larger the average deviation from $\sigma{\bf e}_z$, illustrated by the cone angle in Fig.~\ref{fig:effect}, the higher the magnetic entropy. The spin deviation strongly depends on the effective exchange ($J_{\text{eff}}$) in each magnetic layer, which is a superposition of the \textit{intralayer} exchange ($J$) and \textit{interlayer} exchange ($J_{\text{ie}}$): $J_{\text{eff}} = J + J_{\text{ie}}({\bf n}\cdot{\bf n}_{\text{ie}})$, where ${\bf n}$ and ${\bf n}_{\text{ie}}$ are the unit vectors describing the orientation of the intra- and inter-layer exchange fields. Since ${\bf n}\uparrow\uparrow{\bf n}_{\text{ie}}$ and ${\bf n}\uparrow\downarrow{\bf n}_{\text{ie}}$ correspond respectively to the AP and P configurations, the field-driven AP-to-P switching leads to a change in $J_{\text{eff}}$, which is large if $J$ and $J_{\text{ie}}$ are comparable. Thus, taking into account that the interlayer RKKY is antiferromagnetic, $J_{\text{eff}}^{\text{AP}} = (J + J_{\text{ie}})$ can be interpreted as RKKY-mediated exchange bias that stabilizes the magnetization of each layer against thermal fluctuations and thereby decreases $S$. In contrast, $J_{\text{eff}}^P = (J – J_{\text{ie}})$ has the opposite effect of RKKY-demagnetizing, increasing $S$.

Typically, $J$ is much stronger than $J_{\text{ie}}$ so the RKKY-demagnetizing effect is negligible. The focus of this work, however, is to design a magnetic multilayer with the intra- and interlayer exchange such as to maximally increase the $J_{\text{ie}}$-to-$J$ ratio, while maintaining reliable AP-to-P switching. We choose Fe/Cr known to have very strong RKKY and dilute the ferromagnetic layers (Fe) with Cr, which significantly reduces the intralayer exchange ($J$). We further choose the thickness of the Cr spacers to correspond to the strongest AP-RKKY coupling, thus producing a greatly increased $J_{\text{ie}}/J$.

A series of [Fe$_x$Cr$_{100–x}$(1.2~nm)/Cr(1.2~nm)]$_8$ multilayers, $x = 40 - 55$~at. \% Fe, was fabricated using dc magnetron sputtering on intrinsic Si (100) substrates~\cite{suppl}. The Cr thickness of $1.2$~nm was found to correspond to the strongest AFM interlayer exchange coupling (1$^\text{st}$ AFM RKKY peak). The relatively high AFM-RKKY saturation fields ($\sim$0.5~T) indicate \AA-smooth interfaces~\cite{kelly94,colino96}. The thickness of the FeCr layers was set to 1.2~nm as a compromise between obtaining a lower AP/P switching field (Zeeman energy is higher for thicker FM layers) and a stronger RKKY-demagnetization (surface effect, stronger for thinner FM’s). As the interface roughness increases with the number of bilayer repetitions and affects the RKKY interaction of the top layers, we have found the optimal number of the FeCr/Cr bilayers to be $N \leq 8$, for which this roughness effect is negligible. 

The magnetic measurements were performed using a VSM in the temperature range of 140–430~K. The ferromagnetic contribution was obtained by subtracting the diamagnetic background from the VSM holder and the Si substrates. The data were normalized by the volume of the FeCr layers, estimated using the nominal thickness and the measured area of the samples.

Multilayers with $x = 45$ and $50$\%, hereafter Fe$_{45}$Cr$_{55}$/Cr and Fe$_{50}$Cr$_{50}$/Cr, have their Curie transition in the FeCr layers conveniently within the measurement range and are the focus of this work.  Additionally, reference multilayers with no RKKY coupling (with the Cr spacer thickness of 3~nm) were fabricated and measured in the same sequence as the focus series.

\begin{figure}[h]
\includegraphics{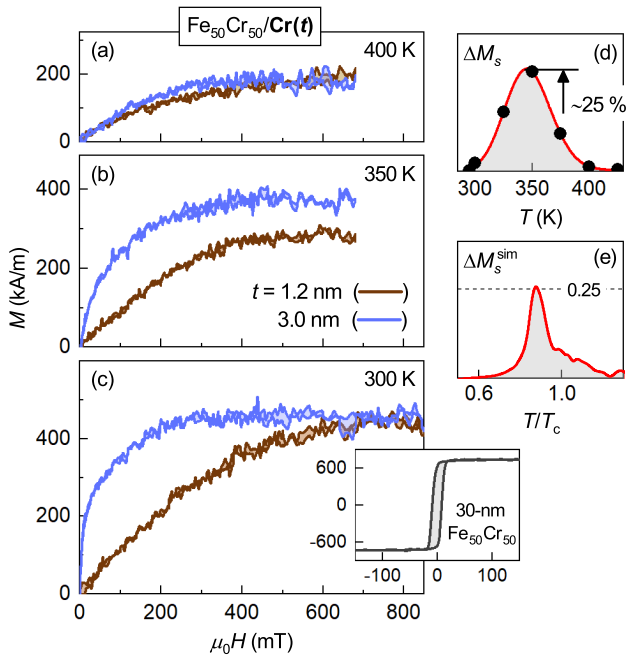}
\caption{\label{fig:VSMcurves} (a)-(c) Magnetization curves for Fe$_{50}$Cr$_{50}$/Cr multilayers with Cr spacer thickness of 1.2 and 3.0~nm for select temperatures. Experimental (d) and spin-atomistically simulated (e) difference in saturation magnetization (at 800~mT) of these two structures versus temperature.} 
\end{figure}

Figure~\ref{fig:VSMcurves} compares the magnetization curves for the multilayers Fe$_{50}$Cr$_{50}$/Cr with the Cr thickness of 1.2~nm (focus sample) and 3.0~nm (reference sample), shown for select temperatures. At room temperature, which is relatively far from $T_{\text{C}} \geq 430$~K of the Fe$_{50}$Cr$_{50}$ layers, the corresponding $M$-vs-$H$ curves are distinctly different. For the focus sample, the observed zero remanence and high saturation field ($\approx700$~mT) clearly indicate a strong AFM-RKKY interlayer coupling. In contrast, the reference sample’s $M$-vs-$H$ is typical for thin ferromagnetic films under strong thermal agitation, with a finite remanence, almost zero coercivity, and a rounded and relatively extended approach to saturation. Thus, the ferromagnetism of the 1.2-nm Fe$_{50}$Cr$_{50}$ layers is strongly affected by the finite-size effects. This is evidenced, in particular, by the measured saturation magnetization of $M_s\approx 4.5\times10^5$~A/m, which is about 60\% of the respective bulk value obtained on a 30~nm thick single film of Fe$_{50}$Cr$_{50}$ [$M_s\approx 7.5\times10^5$~A/m at 300~K, inset to Fig.~\ref{fig:VSMcurves}(c)].

The $M_s$ values for the focus and reference samples are equal at room temperature (the $M$-$H$ curves merge at high fields), which indicates that the FM-layers’ intrinsic exchange is much stronger than the interlayer RKKY exchange at this temperature -- the RKKY in the 1.2-nm sample has no effect on the saturation in the P-state of the multilayer. With increasing temperature toward the Curie point of the dilute ferromagnetic layers, $M_s$ for the focus structure is reduced by up to 25\% compared to $M_s$ for the reference structure [Figs.~\ref{fig:VSMcurves}(b),(d)]. This $\Delta M_s$ is a direct measure of the demagnetizing effect of switching the $J$-vs-$J_{\text{ie}}$ orientation (under AP-to-P switching) on the magnetic entropy of the system. The fact that the strong RKKY-induced demagnetization is observed in a relatively narrow temperature interval below $T_{\text{C}}$ ($300~\text{K} < T < 375~\text{K}$ in this case of Fe$_{50}$Cr$_{50}$/Cr) is due to the nonlinear temperature dependence of the $J_{\text{ie}}/J$ ratio discussed below.

\begin{figure*}[hbt!]
\includegraphics{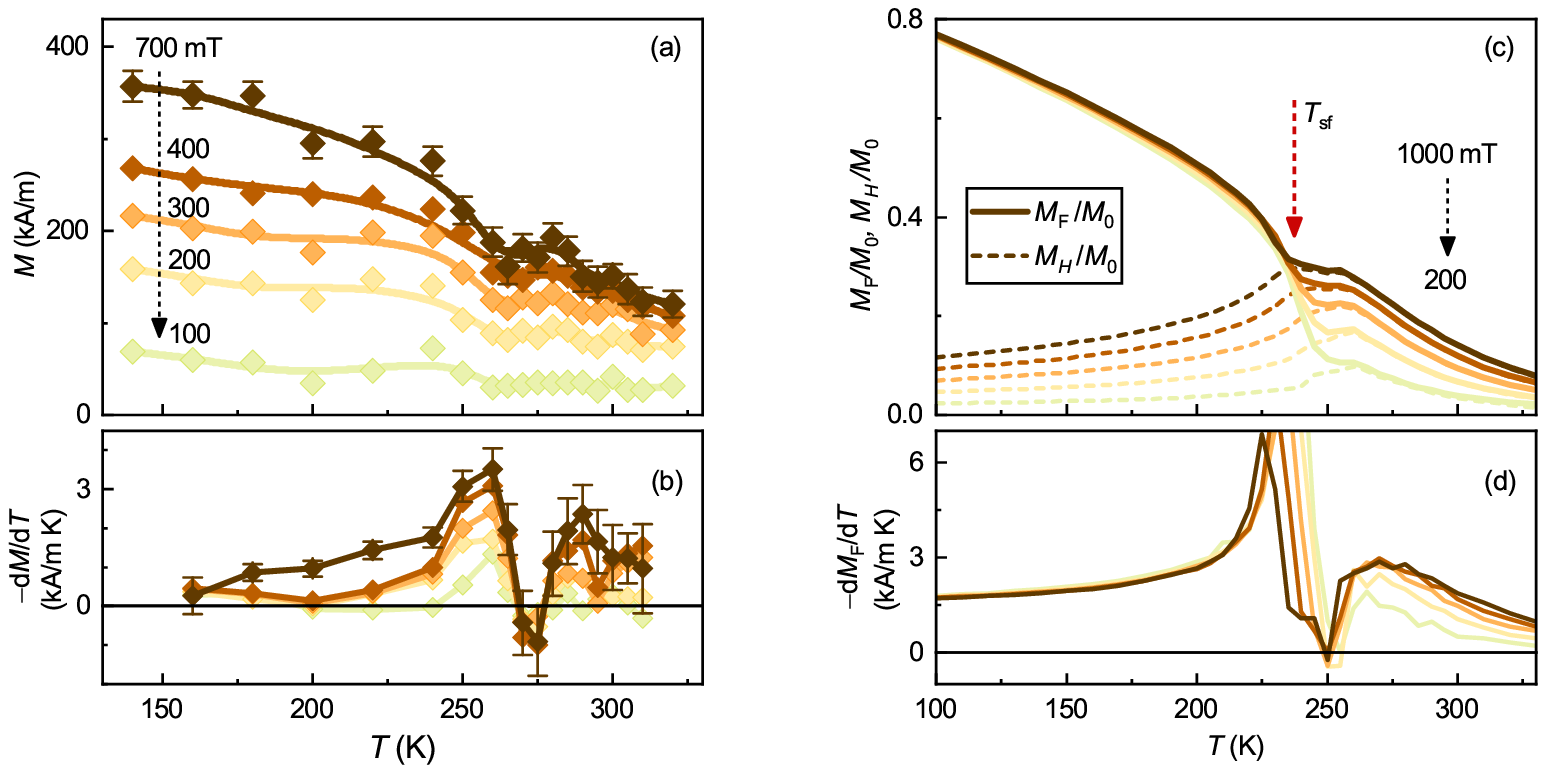}
\caption{\label{fig:MvT} (a),(b) Measured $M$ and d$M$/d$T$ versus $T$ for RKKY-coupled Fe$_{45}$Cr$_{55}$/Cr, $t_{\text{Cr}} = 1.2$, at different magnetic fields, $\mu_0H = 100–700$~mT. (c),(d) Simulated F-layer magnetization, $M_{\text{F}}$, and its derivative, d$M_{\text{F}}$/d$T$, using atomistic spin dynamics~\cite{evans14}. Dashed lines in (c) show calculated magnetization projections, $M_H$, onto direction of applied magnetic field $H$.}
\end{figure*}

\begin{figure}[h]
\includegraphics{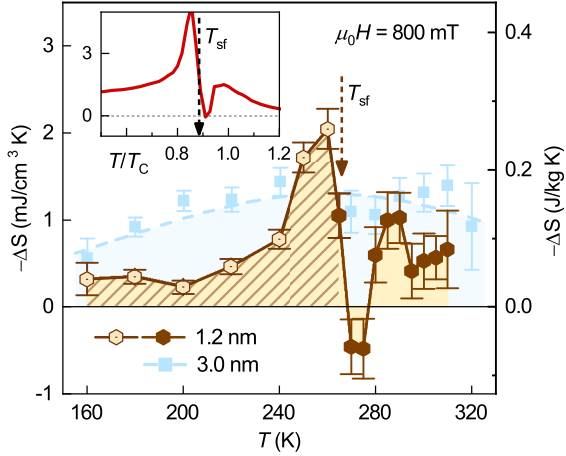}
\caption{\label{fig:entropyChange} Isothermal entropy change versus temperature in field interval 0–800~mT for Fe$_{45}$Cr$_{55}$/Cr multilayers with $t_{\text{Cr}} = 1.2$ and 3.0~nm; data in spin-flop state (below $T_{\text{sf}}$) are shown with open symbols. Inset shows entropy change obtained from atomistic-spin simulated d$M_{\text{F}}$/d$T$ of Fig.~\ref{fig:MvT}(d).}
\end{figure}

We next analyze the RKKY-induced demagnetization and the associated magnetocaloric effect for the Fe$_{45}$Cr$_{55}$/Cr system with strong RKKY ($t_{\text{Cr}} = 1.2$~nm), which has a somewhat lower $T_{\text{C}}=350$~K, extracting the isothermal entropy change from the measured data using the Maxwell relation~\footnote{This procedure is fully valid at $T > T_\text{sf}$, above the 'scissor-state' region, where the magnetic layers are aligned in parallel, and where the inverse MCE actually is observed.}. The $M$-vs-$H$ curves, measured at different temperatures, were translated into $M$-vs-$T$ at various applied field values. Figure~\ref{fig:MvT}(a) shows a pronounced change in $M$-vs-$T$ as the applied field is varied, while the reference sample ($t_{\text{Cr}} = 3.0$~nm; not shown) is essentially field-insensitive in the same range. Moreover, $M$-vs-$T$ for the focus sample shows a step-like decrease at 250–280~K, which is particularly clear in the temperature derivative, $\text{d}M/\text{d}T$ [Fig.~\ref{fig:MvT}(b)]. This non-linear variation is entirely absent in the data for the RKKY-free sample (Fig.~\ref{fig:entropyChange}, $t_{\text{Cr}} = 3.0$~nm) and is a clear signature of the RKKY-driven thermo-magnetic disorder within the F-layers. 

The isothermal entropy change, $\Delta S$, was calculated by field-integrating (d$M$/d$T)|_H$ for each temperature using the Maxwell relation, $\Delta S|_T = –\int\text{d}H(\text{d}M/\text{d}T)|_H$ (Fig.~\ref{fig:entropyChange}). For the focus Fe$_{45}$Cr$_{55}$/Cr structure, the obtained $\Delta S$-vs-$T$ dependence shows a sign reversal in the temperature interval 260–280~K. The negative values of $-\Delta S$ at $T \approx 260–270$~K in Fig.~\ref{fig:entropyChange} mean that the inverse MCE is dominating and the RKKY-demagnetizing effect is most pronounced here, in competition with the normal MCE observed at all other temperatures. As expected, $\Delta S$-vs-$T$ for the reference sample (3.0~nm spacer; blue) is smooth, with $-\Delta S$ positive at all temperatures, which is the most common, \textit{normal} MCE.

The experimental findings can be described in terms of a microscopic mean-field model explaining the origin of the effect and a semiquantitative Landau model addressing the temperature changes in the MCE near $T_{\text{C}}$. The models are physically largely equivalent but emphasize different aspects of the physics involved.

The microscopic model of our system is based on the Hamiltonian $\mathcal{H} = – 2 \mu_0\mu_{\text{B}}{\bf H}_{\text{eff}}\cdot{\bm\sigma}$, where $H_{\text{eff}} = H + \lambda({\bf n} J + {\bf n}_{\text{ie}} J_{\text{ie}})$. Here ${\bf H} = H{\bf e}_z$ is the external magnetic field, $\lambda$ is the molecular-field constant, $J$ is the interatomic exchange inside each layer, and $J_{\text{ie}}$ describes the net RKKY interaction between the layers. The unit vectors ${\bf n} = \pm {\bf e}_z$ and ${\bf n}_{\text{ie}} = \pm {\bf e}_z$ describe the direction of the interatomic and RKKY exchange fields, respectively.

Assuming small magnetization angles, $|{\bf m}_{\perp}^2| = |m_x^2 + m_y^2| \ll 1$, such approach yields the harmonic Hamiltonian $\mathcal{H}= \frac{1}{2}h_{\text{eff}} m_{\perp}^2$, where $h_{\text{eff}} = 2\mu_0\mu_{\text{B}}\sigma{\bf H}_{\text{eff}}\cdot{\bf n}$ is the effective field measured in energy units:
\begin{equation}
    h_{\text{eff}} = 2\mu_0\mu_{\text{B}}\sigma[{\bf H}\cdot{\bf n} + \lambda(J + {\bf n}\cdot{\bf n}_{\text{ie}} J_{\text{ie}})].
\end{equation}
Here, a positive $h = 2\mu_0\mu_{\text{B}}\sigma H$, which is an energy equivalent of the external magnetic field, favors FM spin alignment (small $m_{\perp}$). The $J$-containing
interatomic-exchange term is always positive as it acts to align the spins along the symmetry axis ($z$-axis). On the other hand, the sign of the RKKY contribution depends on the sign of
$J_{\text{ie}}$ (negative in the ground state of zero field in the present experiment) and on the product ${\bf n}\cdot{\bf n}_{\text{ie}}$. This product is respectively positive and negative for AP and P interlayer configurations. 

The entropy $S = -\partial F/\partial T$ is obtained from the free energy $F = -k_{\text{B}}T
\ln Z$, where the partition function is $Z = k_{\text{B}}T/2\pi h_{\text{eff}}$. Since
$h_{\text{eff}}$ depends on the magnetic orientation of the two layers, the result is different entropy in the P and AP states~\cite{suppl}: 
\begin{equation}
    S_P = S_0 - \ln\big[\lambda (J - |J_{\text{ie}}|)/H_T\big] - H/\big[\lambda (J - |J_{\text{ie}}|)\big],
    \label{eqn:Sp}
\end{equation}

\begin{equation}
    S_{AP} = S_0 - \ln\big[\lambda (J + |J_{\text{ie}}|)/H_T\big] +\frac{1}{2}H^2/\big[\lambda^2 (J + |J_{\text{ie}}|)^2\big],
    \label{eqn:Sap}
\end{equation}
where parameter $H_T=k_B T / 2 \pi \mu_0 \mu_B \sigma$ expresses the strength of thermal disorder in units of field (independent of $H$). This explains the experimentally observed negative MCE once the external field switches the spin configuration from AP to P, as illustrated in Fig. 5(a).

Near the Curie temperature ($T_{\text{C}}$), $m_{\perp}$ is no longer small and anharmonic effects with respect to $m_x$ and $m_y$ become important. As $T_{\text{C}}$ is approached from below ($T < T_{\text{C}}$), it is more convenient to apply the Landau expansion for the energy per atom:
\begin{equation}
    F = \frac{1}{2} k_{\text{B}} (T - T_{\text{C}}) m_z^2 + \frac{1}{4}k_{\text{B}} T_{\text{C}}
    m_z^4 - h m_z .
    \label{eqn:freeEnergy}
\end{equation}
Then, the model's entropy $S = - \partial F/\partial T$ is equal $S = - k_{\text{B}} m_z^2$. In zero field, $\partial F/\partial m_z = 0$ yields the spontaneous magnetization $m_z = m_0$, where $m_0 = \pm \varepsilon_{P,AP} (1 - T/T_{\text{C}})^{1/2}$. Here $\varepsilon_{P,AP}$ are the RKKY demagnetizing ($\varepsilon_P$) and biasing ($\varepsilon_{AP}$) factors corresponding to the P and AP configurations, respectively; $\varepsilon_P \leq \varepsilon_{AP}$. At the same time, the effect of small external field ($h\ll k_{\text{B}}T_{\text{C}}$) depends on the sign of $m_0$. By substituting $m_z = m_0 + \mu$ into Eq.~\eqref{eqn:freeEnergy}, doing a series expansion for small $\mu$, and minimizing with respect to $\mu$, one obtains
\begin{equation}
    S_P = S_0 - k_{\text{B}}\varepsilon_P^2(1 - T/T_{\text{C}})-\frac{h}{2T_{\text{C}}\varepsilon_P\sqrt{1
    - T/T_{\text{C}}}},
    \label{eqn:SpFinal}
\end{equation}
\begin{equation}
    S_{AP} = S_0 - k_{\text{B}}\varepsilon_{AP}^2(1 -
    T/T_{\text{C}})+\frac{h^2}{4k_{\text{B}}T_{\text{C}}^2\varepsilon_{AP}^4(1
    - T/T_{\text{C}})^2}.
    \label{eqn:SapFinal}
\end{equation}

\begin{figure}[h]
\includegraphics{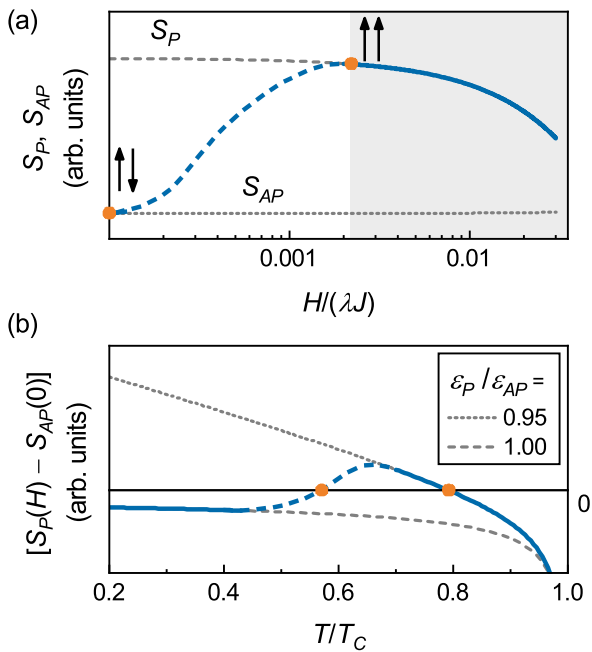}
\caption{\label{fig:switching} 
(a) Magnetic entropy change due to AP-P switching from Eqs.~(\ref{eqn:Sp},\ref{eqn:Sap}). Thick dash line is  guide-to-eye in reorientation interval with noncollinear FM-layers’ magnetic moments. (b) Entropy difference between P and AP configurations versus temperature for $\varepsilon_P/\varepsilon_{AP}$ ratio of 0.95 and 1.00, from Eqs.~(\ref{eqn:SpFinal},\ref{eqn:SapFinal}). Thick dash line illustrates transition between two regimes due to temperature-induced change in $\varepsilon_P/\varepsilon_{AP}$.}
\end{figure}

Expressions~\eqref{eqn:SpFinal} and \eqref{eqn:SapFinal} show that the system's entropy is dominated by $S_P$ (normal MCE) when the RKKY has a minor effect ($\varepsilon_P\approx\varepsilon_{AP}$), and by $S_{AP}$ (inverse MCE) when the RKKY biasing/demagnetizing becomes stronger ($\varepsilon_P \lesssim \varepsilon_{AP}$). Noting from the experiment that $\varepsilon_P/\varepsilon_{AP}$ varies with temperature, this can explain
the observed temperature-induced variation in the sign of the MCE. Figure~\ref{fig:switching}(b) illustrates, using Eqs.~(\ref{eqn:SpFinal},\ref{eqn:SapFinal}), how the inverse MCE can become the dominating contribution in a finite temperature interval [positive ($S_P - S_{AP}$)].

To verify the above analytical results, we have used VAMPIRE~\cite{evans14} to perform atomistic spin simulations of our Fe$_{50}$Cr$_{50}$/Cr/Fe$_{50}$Cr$_{50}$ system, modelled as a random alloy with a bcc unit cell. The effect of the Cr spacer of thickness corresponding to the first antiferromagnetic RKKY peak (the experimental case) was approximated by a negative interlayer exchange interaction (of AFM type) between the nearest-neighboring Fe-Cr layers of strength $J_{\text{RKKY}} = -0.05J_{\text{int}}$, where $J_{\text{int}}$ is the intrinsic exchange constant in iron, $J_{\text{Fe}} = 6.25\times10^{-21}$~J/atomic-volume. A multilayer was obtained by using periodic boundary conditions~\cite{suppl}. The initial simulation was for $M(T)$, stepping temperature in a constant applied field (200, 400, ..., 1000~mT) with the individual layers’ magnetizations initialized in a `scissor state' (symmetric about the field) and $J_{\text{RKKY}} = 0$. The RKKY interaction strength was then taken to scale with the layer’s magnetization and the simulation was repeated with $J_{\text{RKKY}}(T) = -0.05J_{\text{int}}m(T)$, where $m(T)=M(T)/M_s$ was taken from the preceding $M$-vs-$T$ simulation. We found this iterative process of incorporating the temperature dependence of the RKKY interaction to be accurate in capturing the physics involved and for correctly describing the experimentally observed behavior.

The results of the atomistic-spin simulations for the F-layer magnetization are shown in Fig.~\ref{fig:MvT}(c) and reproduce well the experimental behavior of Fig.~\ref{fig:MvT}(a). The respective temperature derivative shown in Fig.~\ref{fig:MvT}(d) exhibits a sign change in a narrow range below the Curie point, in excellent agreement with the experiment [Fig.~\ref{fig:MvT}(b)]. The simulated data converted to entropy in the inset to Fig.~\ref{fig:entropyChange} as well as to $\Delta M(T)$ in Fig.~\ref{fig:VSMcurves}(e) additionally confirm the measured behavior. The atomistic-spin simulations thus fully agree with the analytical model and the experiment. 

The inversion of MCE observed in our Fe-Cr/Cr multilayers is, at first glance, counter-intuitive as it corresponds to an increase in magnetic disorder when the field is increased toward saturation. Such effect has been discussed theoretically for similar antiferromagnetically coupled Co/Cr superlattices, however, only the normal MCE was observed~\cite{mukherjee09}, likely due to a large $J/J_{\text{ie}}$ ratio for the pure Co FM-layers used. The key in our experiment for inverting the MCE was to optimally dilute the FM-layers and thereby reduce the $J/J_{\text{ie}}$ ratio, and at the same time make sure that the RKKY exchange energy is comparable to the thermal energy in the temperature interval of interest, such that the material could be cycled to and away from its effective $T_C$ by switching the direction of the RKKY term (under AP to P switching).

Our microscopic analytical model as well as numerical atomistic-spin simulations explain the inverse MCE observed for our system as originating from the difference in the layers’ effective exchange $J_{\text{eff}} = J \pm J_{\text{ie}}$ for the AP (``+'') and P (``–'') configurations. By introducing a temperature-dependent parameter $\varepsilon_{P,AP}$ characterizing the RKKY-induced de/magnetizing effect, we show how the negative MCE can dominate in a finite temperature interval near $T_{\text{C}}$ – the behavior observed on the experiment [Figs.~\ref{fig:VSMcurves},\ref{fig:entropyChange},\ref{fig:switching}(b)]. Calculating the exact functional form of $\varepsilon_P/\varepsilon_{AP}(T)$ analytically is non-trivial and goes beyond the scope of this paper.

The demonstrated inverse MCE in RKKY-coupled multilayers has the potential for further enhancements via, e.g., increasing the $J_{\text{ie}}$-to-$J$ ratio by high-polarization magnetic doping at the F/N interfaces~\cite{parkin92,dpol-apl20}. The ideal case would be full spin disordering in the structure on RKKY switching. The effect should be useful for micro-coolers and heat-exchangers, as well as spin-thermionic applications such as heat-assisted memory.

\begin{acknowledgments}
This work has benefited from discussions with Christian Binek. Support by the Swedish Research Council (VR\#2018-03526), the Olle Engkvist Foundation (2020-207-0460) and, in Nebraska, by NSF EQUATE (OIA-2044049) and NCMN are gratefully acknowledged.
\end{acknowledgments}

\bibliography{invMceBib}

\end{document}